\documentclass[aps,pra,showpacs,amsmath,amssymb,amsfonts,twocolumn,superscriptaddress,floatfix,nofootinbib]{revtex4} 
\usepackage{psfig}
\usepackage{graphicx}

\newcommand{\eqname}[1]{\label{eq:#1}}
\newcommand{\bgar}{\begin{eqnarray}}
\newcommand{\enar}[1]{\label{eq:#1}\end{eqnarray}}
\newcommand{\valass}[1]{\left|#1\right|}

\newcommand{\kk}{ {\bf k}}

\newcommand{\vel}{ {\bf v}}

\newcommand{\eps}{\epsilon}

\newcommand{\eq}[1]{(\ref{eq:#1})}
\newcommand{\al}[1]{^{(#1)}}

\begin{document}

\title{Transverse Fresnel-Fizeau drag effects in strongly dispersive media.}

\affiliation{Laboratoire Kastler Brossel, \'Ecole Normale Sup\'erieure, 24
rue Lhomond, 75231 Paris Cedex 05,
France} 

\affiliation{INFM, Department of Chemistry and Physics of
Materials, Via Valotti 9, 25133 Brescia, Italy}

\affiliation{INFM, European Laboratory for non-Linear Spectroscopy, Via N.
Carrara 1, 50019 Sesto Fiorentino, Italy.} 

\affiliation{Scuola Normale Superiore and INFM, Piazza dei Cavalieri 7,
I-56126 Pisa, Italy}

\author{I. Carusotto}
\email{Iacopo.Carusotto@lkb.ens.fr}
\affiliation{Laboratoire Kastler Brossel, \'Ecole Normale
Sup\'erieure, 24 rue Lhomond, 75231 Paris Cedex 05, France}

\author{M. Artoni}
\affiliation{INFM, Department of Chemistry and Physics of Materials, Via
Valotti 9, 25133 Brescia, Italy}
\affiliation{INFM, European Laboratory for non-Linear Spectroscopy, Via N.
Carrara 1, 50019 Sesto Fiorentino,
Italy.}

\author{G. C. La Rocca}
\affiliation{Scuola Normale Superiore and INFM, Piazza dei Cavalieri
7, I-56126 Pisa, Italy}

\author{F. Bassani}

\affiliation{Scuola Normale Superiore and INFM, Piazza dei Cavalieri 7, I-56126
Pisa, Italy}

\begin{abstract}

A light beam normally incident upon an uniformly moving dielectric medium is in
general subject to bendings due to a transverse Fresnel-Fizeau light drag
effect. In conventional dielectrics, the magnitude of this bending effect is
very small and hard to detect.  Yet, it can be dramatically enhanced in 
strongly dispersive media where slow group velocities in the m/s range have
been recently observed taking advantage of the electromagnetically induced
transparency (EIT) effect. 
In addition to the usual downstream drag that 
takes place for positive group velocities, we predict a significant anomalous
upstream drag to occur for small and negative group velocities.
Furthermore, for sufficiently fast speeds of the medium, higher order
dispersion terms are found to play an important role and to be responsible for
peculiar effects such as light propagation along curved paths and the
restoration of the spatial coherence of an incident noisy beam. The physics
underlying this new class of slow-light effects is thoroughly discussed.

\end{abstract}




\date{\today}

\maketitle

\section{Introduction}

\label{sec:Intro}

A constant effort has always be devoted to the search for
new effects and materials to control the propagation of
light waves. Over the past few years, in particular, the use of quantum
interference has 
led to an astonishing control of light
waves propagating through specific classes of atomic and solid state media.
These materials exhibit superior
properties that cannot be found in conventional ones. 
{\em Slow light} propagation at group velocities as small as 1 m/s, e.g., has
been observed in experiments with 
Bose-Einstein condensates of sodium atoms~\cite{SlowLightColdAtoms-hau,SlowLightColdAtoms-inouye},
in hot rubidium vapors~\cite{SlowLightHotAtoms-kash,SlowLightHotAtoms-budker} as well as in solid
doped Pr:Y$_2$SiO$_5$~\cite{SlowLightSolids1} crystals and
Ruby~\cite{SlowLightSolids2}.
Reversible {\em stopping} of a laser
pulse~\cite{LuceFermaTh} in ultracold and hot alkali
vapors~\cite{LuceFermaExp-phillips,LuceFermaExp-hau} as well as in Pr:Y$_2$SiO$_5$ crystal has
also been observed~\cite{SlowLightSolids1}.
 
Light stopping and slow-light propagation effects originate from
electromagnetically induced transparency (EIT). Such a widely discussed
phenomenon arises from quantum interference and is characterized by a strong
enhancement of the refractive index dispersion within a narrow frequency
window around the medium resonance where absorption turns out to be largely
quenched.
The study of such a phenomenon goes back to the late seventies when 
non-absorbing resonances in atomic sodium have first been observed by
Gozzini's group and later interpreted in terms of coherent
population trapping~\cite{Alzetta,EITReviewArimondo,Hansch}.

The interest for such a phenomenon has revived~\cite{HarrisReview} over the
past decade and has now become a rather topical area of
research~\cite{SlowLightReview,Marangos}, much work being done
on fundamental issues: high nonlinear coupling between weak
fields and quantum entanglement of slow
photons~~\cite{NLO-kasapi,NLO-harris,NLO-lukin}, 
entanglement of
atomic ensembles~\cite{AtomEntangl}, quantum memories~\cite{QuantumMem}
and enhanced acousto-optical effects~\cite{AcoustoOpt}, just to mention a
few. In particular, a strict analogy
between slow-light in moving media and light propagating in curved
space-times has been unveiled and some of its
consequences have been recently
discussed~\cite{ulf-prl-2000,ulf-pra-2000,ulf-nature,ulf-black-hole,ulf-pulse,ulf-jmo-2001,ulf-physicsworld}.

Recently it has been also anticipated in~\cite{ArtoniNegVg} that EIT media
are ideal candidates for the observation of extremely low and negative group
velocities and (apparently) superluminal behaviour. 
In such media, the group velocity can in fact be
readily tuned over a wide range of negative values directly by varying the
coupling and probe detunings in a standard three-level
 $\Lambda$ configuration. 

Although most slow light and negative group velocity experiments have dealt
with the basic problem of a light 
pulse delay during its propagation across the dispersive medium,
ultraslow positive or negative group velocities can have interesting
consequences in many different scenarios. In this paper we present a thorough
investigation of slow-light propagation through a moving medium.
Specifically, we examine a configuration in which a highly dispersive EIT 
medium which moves with uniform velocity and normally
\footnote{A
  configuration in which the medium is set to move parallel to the probe 
light-beam, leading to the more familiar
{\em longitudinal} Fresnel-Fizeau effect, has been
studied in~\cite{FresLongEIT}. 
Also in this configuration, the magnitude of the light-drag effect is
predicted to be strongly enhanced in the slow light regime.} 
to an incident light beam of finite spatial extent. The same
geometry was adopted in the early seventies by Jones in his pioneering work
on the {\em transverse} Fresnel-Fizeau light
drag effect~\cite{Jones,Jones2} leading to the observation of a very small
downstream bending, i.e., in the direction of motion, of a light ray. 
The use of a strongly dispersive
medium supporting slow light, rather than the
non-dispersive glassy material used by Jones, will not only
allow for a remarkable enhancement of the drag effects but also for
qualitatively new features.

The possibility of having light propagating with small negative
group velocities across the dragging
medium is predicted to yield large {\em upstream} light bendings,
i.e. in the direction opposite to that of the medium. The phenomenology of
such an a anomalous Fresnel-Fizeau light drag effect, which has been the
subject of an old controversy during the late
seventies~\cite{KoChuang,Lerche}, is here discussed in detail unwinding some
of its controversial aspects.

For sufficiently fast dragging speeds, the correct description of
the slow light transverse dragging effect requires that absorption dispersion
and group velocity dispersion be taken into account. As we shall see below, 
it turns out that these higher--order dispersion terms introduce new and
peculiar features, such as propagation along curved light paths and the
restoration of the spatial coherence of a noisy beam.
All the numerical results presented
in this paper have been obtained using realistic parameters taken from
slow-light experiments in ultra-cold atomic
clouds~\cite{SlowLightColdAtoms-hau}. Since the physics 
of the system is essentially determined by the electromagnetically induced
transparency effect, the physical features here discussed are however extremely
general and hold through also for the recently prepared solid state EIT media.

The paper is organized as follows. In sec. \ref{sec:GeneralTh} we introduce 
the physical system and we present the model used for our
predictions. The general theory of the transverse
Fresnel-Fizeau light drag effect is presented in sec. \ref{sec:Downstream}. 
These general results are specialized to
the case of a strongly dispersive dressed medium driven into a $\Lambda$
configuration by a resonant and non-resonant
coupling beam respectively in sec. \ref{sec:DownstreamEIT} and in sec.
\ref{sec:Upstream}: in the former case, the group velocity is small and
positive, so the transverse Fresnel drag occurs in the usual downstream
direction; in the latter case, the negative group velocity is shown to give
an anomalous upstream Fresnel drag.
In sec.\ref{sec:HigherOrder} we proceed to discuss some effects which
arise from the inclusion of higher order dispersion terms.
We carry out a complete analysis for two specific instances. In one case,
the beam spectrum reshaping due to
absorption and group velocity dispersion, which makes the average group
velocity to bend during propagation, is seen to lead to curved light
paths. In the other, the frequency dispersion of
absorption is seen to act as a filter for spatial fluctuations of the beam,
whose space coherence properties improve as it propagates through the
medium.
Finally, a summary of the work is given in sec.\ref{sec:Conclu} where
conclusions are also drawn.

\section{The model and general theory}

\label{sec:GeneralTh}
\begin{figure}[htbp]
\includegraphics[width=3.3in,clip]{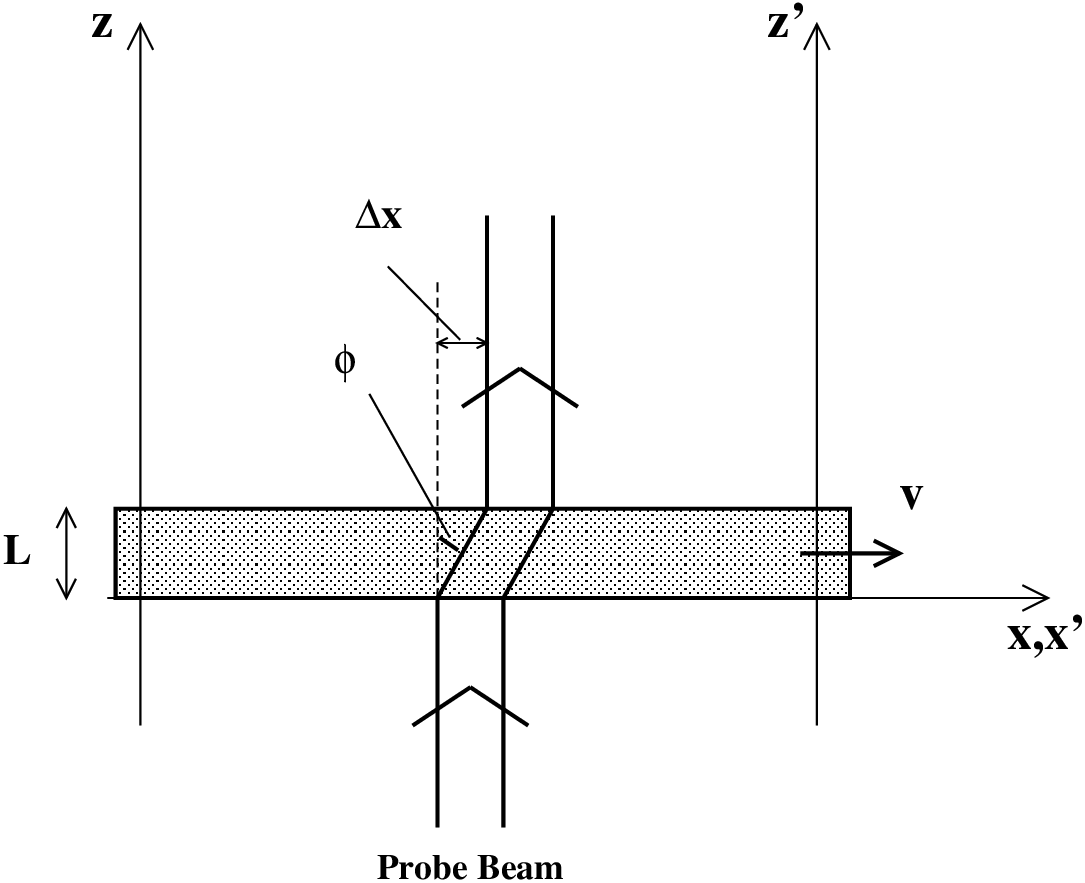}
\caption{
Scheme of the experimental set-up under
consideration.
\label{fig:Setup}}
\end{figure}

We consider a monochromatic {\em probe} light beam of frequency $\omega_0$
 propagating along the $z$ axis and normally incident upon a homogeneous
 dielectric medium uniformly moving along the $x$ direction as shown in
 fig.\ref{fig:Setup}. We here denote with $L$ the thickness of the slab and
 with $v$ its velocity $v\ll c$.
The probe beam
has a Gaussian profile centered
 in $(x_0,y_0)$ so that at $z=0$ one has, 
\begin{equation}
E_0(x,y)=E_0\,e^{-[(x-x_0)^2+(y-y_0)^2]/2\sigma_0^2},
\eqname{Waist}
\end{equation}
where $\sigma_0$ is the beam waist.
The corresponding Fourier transform ${\tilde E}_0(k_x,k_y)$ of $E_0(x,y)$ is
then a Gaussian of width $\sigma_0^{-1}$.

In the following we shall always restrict our attention to the case of
a weak probe beam, which allows us to apply linear response
theory and describe the polarization of the medium
by means of a dielectric function $\epsilon$. Assuming
for simplicity a non-magnetic ($\mu=1$) and isotropic
$\epsilon_{i,j}=\epsilon\,\delta_{i,j}$ medium, the scalar dielectric function 
$\epsilon$  completely characterizes the
linear polarization of the medium in its rest frame $\Sigma'$.
Spatial locality of the dielectric
polarization will also be assumed, i.e.
$\epsilon$ will be taken to depend on the
frequency $\omega'$ but not on the wave-vector $\kk'$.
Primed quantities will refer to the
medium rest frame $\Sigma'$, while the non-primed ones will refer to the
laboratory frame $\Sigma$.
The linearity of the optical response and the translational
invariance of the system on the $(x,y)$ plane allow us to 
decompose the
incident field in its Fourier components in the $(x,y)$ plane, to
propagate each of them independently from the others, and to finally
reconstruct
the transmitted beam profile by using an inverse Fourier
transform.

In the $\Sigma'$ frame, the dispersion law inside the medium has the
usual form:
\begin{equation}
\epsilon(\omega')\, {\omega'}^2=c^2 ({k_x'}^2 +{k_y'}^2 +{k_z'}^2).
\eqname{FresnelWave'}
\end{equation}
Since the slab speed $v\ll c$,
the linearized form of the Lorentz transformations can be used,
namely:
\begin{eqnarray}
\omega'&=&\omega-k_x v \label{eq:Lorentz1} \\
k_x'&=&k_x-\frac{\omega}{c^2}v \label{eq:Lorentz2} \\
k_{y,z}'&=&k_{y,z}. \label{eq:Lorentz3}
\end{eqnarray}
Notice that even though the medium does exhibits no spatial dispersion in its
rest frame $\Sigma'$, spatial dispersion arises however in the laboratory
frame $\Sigma$ from the dependence of $\omega'$ on $k_x$ 
\footnote{In the rest frame $\Sigma'$, the dielectric polarization of a
  medium with frequency dispersion depends on the retarded values of
the electric field at the same  spatial position.
As seen from the laboratory frame $\Sigma$, the polarization at a given point
will thus depend on the electric field at different spatial positions, which
means that a moving medium shows a spatial dispersion in $\Sigma$ even if it
is only temporally dispersive in the rest frame $\Sigma'$.}.
By inserting the transformations (\ref{eq:Lorentz1}-\ref{eq:Lorentz3})
into the dispersion law \eq{FresnelWave'}, one obtains the following expression:
\begin{equation}
\frac{(\omega-k_x v)^2}{c^2}\,\epsilon(\omega-k_x v)=\big(k_x-\frac{\omega
v}{c^2}\big)^2 +k_y^2 +k_z^2.
\eqname{FresnelWave}
\end{equation}

This equation represents the dispersion law
in $\Sigma$ and can be used to determine the propagation of the light
beam across the slab.
Knowing the frequency $\omega_0$ and the transverse components
$k_{x,y}$ of the wavevector allows us to obtain from \eq{FresnelWave} the
$z$ component $k\al{in}_z$ of the wave-vector inside the medium:
\begin{multline}
k_z\al{in}(k_x,k_y)=\\=\sqrt{\big(\frac{\omega_0-k_x
v}{c}\big)^2\,\epsilon(\omega_0-k_x v)-\big(k_x-\frac{\omega_0
v}{c^2}\big)^2-k_y^2},
\eqname{k_zeps}
\end{multline}
as well as in the external free space:
\begin{equation}
\eqname{k_z0}
k_z\al{out}(k_x,k_y)=\sqrt{\frac{\omega_0^2}{c^2}-k_x^2-k_y^2}.
\end{equation}

The profile of the transmitted beam can then be obtained
by decomposing the field in its Fourier components at transverse
wavevector $(k_x,k_y)$ and propagating each of them independently with the
wave-vectors \eq{k_zeps} and \eq{k_z0}.
Since the energy flux must be along the positive $z$ axis,
the roots with positive real part
$\textrm{Re}[k_z\al{in,out}]>0$ have to be taken
\footnote{As the reflection amplitude is proportional to $\epsilon-1$, the
  approximation of neglecting interface reflections is a reasonable
  approximation in the case of an ultra cold atomic gas on
  which the present paper is focussed. On the other hand, a more complete
  theory including the possibility of multiple interface reflections is
  required in the case of a solid state medium.}
.

For each component, the amplitude at the position $0\leq z\leq L$ inside the
slab is given by:
\begin{equation}
{\tilde E}(k_x,k_y;z)=e^{i \Phi(k_x,k_y;z)}{\tilde E}_0(k_x,k_y)
\eqname{Amplitude_z}
\end{equation}
with a phase $\Phi$:
\begin{equation}
\Phi(k_x,k_y;z)=k\al{in}_z(k_x,k_y)\,z
\end{equation}
Notice that the wavevector $k\al{in}_z(k_x,k_y)$ as well as the phase
$\Phi(k_x,k_y;z)$ are generally complex
quantities; their imaginary parts vanish only for non-absorbing, non-amplifying medium.
Past the slab, the transmitted amplitude at the position $z>L$ is
given by the same equation \eq{Amplitude_z} with the phase $\Phi$ now 
given by:
\begin{equation}
\Phi(k_x,k_y;z)=k\al{in}_z(k_x,k_y)\,L+k\al{out}_z(k_x,k_y)\,(z-L).
\eqname{TransmPhase}
\end{equation}
The spatial profile of the transmitted beam at any point $z$ can then
be obtained taking the inverse Fourier transform of \eq{Amplitude_z},
\begin{equation}
E(x,y,z)=\int\!\frac{dk_x\,dk_y}{2\pi}\,e^{i(k_x x+k_y
y)}\,e^{i\Phi(k_x,k_y;z)}\,{\tilde E}_0(k_x,k_y).
\eqname{InvFourier}
\end{equation}

\section{The transverse Fresnel-Fizeau drag effect}
\label{sec:Downstream}
Provided that the incident beam waist $\sigma_0$ is wide enough, only a very small
window of wave-vectors $(k_x,k_y)$ around $k_{x,y}=0$ is effectively
relevant to the propagation dynamics and one can safely expand the
phase \eq{TransmPhase} in powers of $k_{x,y}$ so that to the lowest order one
has: 
\begin{multline}
\Phi(k_x,k_y;z)=\frac{\omega_0}{c}\sqrt{\epsilon(\omega_0)}\left[1-\frac{k_x
v}{\omega_0}\left(1-\frac{1}{\epsilon(\omega_0)}\right) \right.
\\\left. -\frac{k_x\,
v}{2\,\epsilon(\omega_0)}\frac{d\epsilon}{d\omega}\right]L+\frac{\omega_0}{c
}(z-L).
\eqname{Phi1}
\end{multline}
The position of the center $(x_c,y_c)$ of the resulting wave packet at
a given $z$ can be obtained inserting the expansion \eq{Phi1} into
\eq{InvFourier} and then invoke the so-called {\em stationary-phase}
principle.
This states that the integral in \eq{InvFourier} has its maximum value
at those points $(x_c,y_c)$ for
which constructive interference between the different Fourier
components occurs, that is at those points at which the phase of the integrand
is stationary~\cite{Jackson}:
\begin{equation}
\left.\frac{\partial}{\partial
k_{x,y}}\left(\textrm{Re}[\Phi(k_x,k_y;z)]+k_x x_c+k_y
y_c\right)\right|_{k_x=k_y=0}=0.
\eqname{StatPhase}
\end{equation}
Inserting into \eq{StatPhase} the expression of the phase \eq{Phi1}
and assuming the 
imaginary part $\epsilon_i$ of $\epsilon=\epsilon_r+i\,\epsilon_i$ to be
negligible, one 
finds that in the laboratory frame $\Sigma$ the beam propagates
inside the moving slab at a non-vanishing angle $\theta$ with respect to the
normal $z$ direction given by:
\begin{multline}
\tan\theta=\frac{v}{c}\left[\sqrt{\epsilon_r(\omega_0)}+\frac{\omega_0\,}{2
\sqrt{\epsilon_r(\omega_0)}}\frac{d\epsilon_r}{d\omega}-\frac{1}
{\sqrt{\epsilon_r(\omega_0)}}\right]=\\
=\frac{v}{c}\,\left[\frac{c}{v'_{\rm gr}}-\frac{v'_{\rm ph}}{c}\right].
\eqname{theta}
\end{multline}
Here $v_{\rm gr}'$ and $v_{\rm ph}'$ denote respectively the group
and phase velocities in the medium rest frame $\Sigma'$:
\begin{eqnarray}
v'_{\rm ph}&=&\frac{c}{\sqrt{\epsilon_r(\omega_0)}} \\
v_{\rm gr}'&=&\frac{c}{\sqrt{\epsilon_r(\omega_0)}+
\frac{\omega_0}{2\sqrt{\epsilon_r(\omega_0)}}\frac{d\epsilon_r}{d\omega}}.
\eqname{GroupVel}
\end{eqnarray}
This deflection of the light beam can be interpreted as a {\em
  transverse} Fresnel-Fizeau drag effect, in which the beam of light is dragged by
  the transverse motion of the medium.
After exiting from the rear surface of the slab, the beam
again propagates along the normal direction. Its center, however, turns
out to be laterally shifted by an amount,
\begin{equation}
\Delta x=Lv\left[\frac{1}{v'_{\rm gr}}-\frac{v'_{\rm ph}}{c^2}\right].
\eqname{Deltax}
\end{equation}
along a direction parallel to the medium velocity. This expression is in
agreement with the ones derived by Player and
by Rogers~\cite{FresTrTh-player,FresTrTh-rogers}.

While the group velocity direction inside the moving medium makes a finite
angle $\theta$
with respect to the normal, the transverse $x$ and $y$ 
components of the phase velocity
are always vanishing in the laboratory frame
$\Sigma$ because the light beam is normally incident on the slab and
the transverse wave vector is conserved at the interface.
The non-parallelism of the group and phase velocities arises then 
from the effective spatial dispersion acquired in the laboratory
frame $\Sigma$  by the moving medium~\footnote{The assumed isotropy of the
  medium in its rest frame $\Sigma'$ guarantees that group and phase
  velocities are always parallel in $\Sigma'$.}.
Indeed, one can calculate the group velocity $\vel_{\rm
  gr}=\nabla_\kk \omega(\kk)$ at $k_{x,y}=0$ directly from the dispersion law
\eq{FresnelWave} to obtain for each component:
\begin{eqnarray}
v_{{\rm gr},x}&=&
v\,\left(1-\frac{1}{\epsilon+\frac{\omega}{2}\frac{d\epsilon}{d\omega}}
\right)=v
\left(1-\frac{v'_{\rm gr}\,v'_{\rm ph}}{c^2}\right)
\label{eq:vel_x} \\
v_{{\rm gr},y}&=&0 \label{eq:vel_y} \\
v_{{\rm gr},z}&=&\frac{c}{\sqrt{\epsilon}+\frac{\omega}
{2\sqrt{\epsilon}} \frac{d\epsilon}{d\omega}}=v'_{\rm gr}\label{eq:vel_z}.
\end{eqnarray}
The non-vanishing value of $v_{{\rm gr},x}$ is responsible for the transverse
Fresnel-Fizeau drag. The value of $\tan \theta=v_{{\rm gr},x}/v_{{\rm gr},z}$
obtained from \eq{vel_x} and \eq{vel_z} agrees indeed with \eq{theta}.

Another picture of this drag effect can be obtained by
working in the rest frame $\Sigma'$~\cite{AnomFresTr}. Because of the
aberration of light~\cite{Jackson}, the direction of the incident beam  makes
an angle
$\theta'_{\rm inc}=-v/c$ with the normal
\footnote{Notice that in the rest frame $\Sigma'$, the beam waist appears as
  uniformly translating along the negative $x$ axis. This unfamiliar
  feature however does not affect the argument.}.
In the rest frame $\Sigma'$, the group and phase velocities inside the
medium are parallel to each other and make an angle $\theta'_{\rm
  refr}=-v/(c\sqrt{\epsilon})$ with the normal according to Snell's law.
The modulus of the group velocity in $\Sigma'$ is given by \eq{GroupVel}.
By Lorentz-transforming back the group velocity to the laboratory
frame $\Sigma$, it is easy to check that the same deflection angle as in
\eq{theta} is obtained.

As one can see from the explicit expression \eq{Deltax} the
magnitude of the transverse drag $\Delta x$ is largest for a strongly
dispersive
medium in which $\epsilon$ has a rapid frequency dependence and the group
velocity $v'_{\rm gr}$ results much slower than the vacuum speed of light
$v'_{\rm gr}\ll c$. 

In this  case, Galilean velocity composition laws
can be safely applied and the result \eq{Deltax} simplifies to
\begin{equation}
\Delta x=\frac{Lv}{v'_{\rm gr}},
\end{equation}
which yields a very intuitive interpretation for $\Delta x$ as the
displacement of the medium during the time interval $\Delta t=L/v'_{\rm gr}$
taken by the light to travel across it.

\section{The transverse Fresnel-Fizeau drag effect in a EIT medium}
\label{sec:DownstreamEIT}

The first experimental observation of the transverse Fresnel-Fizeau drag
effect was performed in the mid-1970's by Jones~\cite{Jones} using a
rotating glass disk as moving dielectric medium.
In such a non-dispersive medium the phase and group velocities were of
the order of $c$ ($c/v'_{\rm gr}\approx c/ v'_{\rm ph}\approx 1.5$) and the
limited rotation velocity of the disk at the laser spot
($v\simeq 2\times 10^4\textrm{cm/s}$) limited the
lateral displacement to a distance of the order of $6\,\textrm{nm}$.
However, a clever optical alignment technique allowed not only
to observe the effect, but also to discriminate the 
validity of the result \eq{theta} from other possible expressions~\cite{Jones2}.

\begin{figure}[htbp]
    \includegraphics[width=2in,clip]{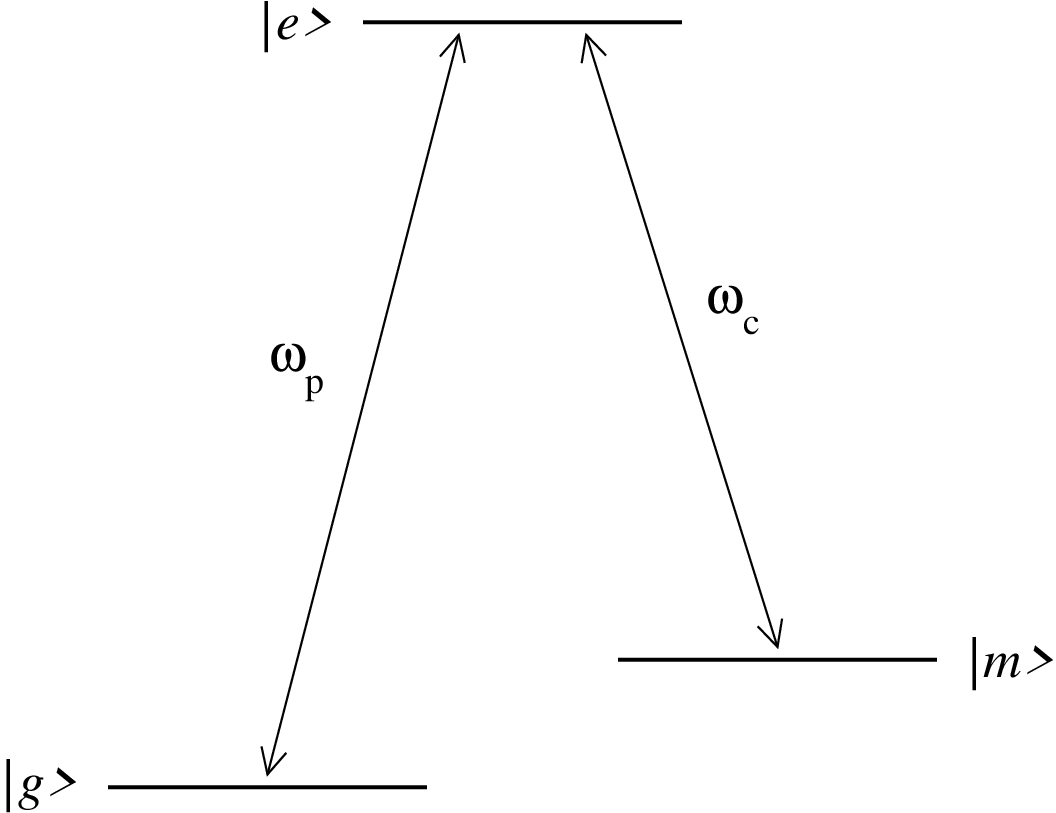}
    \caption{Scheme of the energy levels involved in the optical transitions.
\label{fig:Levels}}
\end{figure}

As remarked in the previous section, the use of a strongly
dispersive medium allows for a significant enhancement of the
drag effect. Very slow group velocities 
can now be obtained in both atomic
samples~\cite{SlowLightHotAtoms-kash,SlowLightHotAtoms-budker,SlowLightColdAtoms-hau,SlowLightColdAtoms-inouye}
and solid-state
media~\cite{SlowLightSolids1,SlowLightSolids2} by optically dressing a
resonant transition with a coherent {\em coupling} laser beam
as shown in the $\Lambda$ level scheme of fig.\ref{fig:Levels}.

Under the assumption of a weak probe beam, linear response theory holds for
the system dressed by the strong coupling beam: the
resulting dielectric constant is to be interpreted as describing the linear
response of the optically driven medium to the weak probe.
Obviously, this dressed dielectric constant has a strongly nonlinear dependence on
the coupling intensity so to account for all the nonlinear processes
induced in the medium by the coupling field.
In the simplest case of non-degenerate levels,  the rest frame dielectric
constant of our optically driven $\Lambda$ configuration acquires the
well-known form~\cite{EITReviewArimondo,HarrisReview,SlowLightReview,Hansch},
\begin{equation}
\epsilon(\omega)=\epsilon_\infty+\frac{4\pi
f}{\omega_e-\omega-i\gamma_e/2-\frac{\valass{\Omega_c}^2}
{\omega_m+\omega_c-\omega-i\gamma_m/2}}
\eqname{EpsEIT}
\end{equation}
where $\omega_{e(m)}$ and $\gamma_{e(m)}$ are respectively the frequency
and the linewidth
of the excited $e$ and metastable $m$ states, where we have set $\omega_g=0$.
Here $\omega_c$ is the frequency of the $e\leftrightarrow m$ coherent
coupling beam
\footnote{Doppler shift of the coupling beam is avoided by choosing
the direction of the coupling beam to be orthogonal to both the probe beam and
the medium velocity. The waist of the coupling beam is taken as much larger than
both the waist $\sigma_0$ of the probe and the thickness $L$ of the medium.}
 and $\Omega_c$ its Rabi frequency.
The linewidth $\gamma_m$ of the metastable state is much smaller than
the linewidth $\gamma_e$ of the excited state.
The $f$ parameter quantifies the oscillator strength of the optical
transition: for an ultra cold atomic
gas~\cite{SlowLightColdAtoms-hau} at atomic densities $n\approx
10^{12}\,\textrm{cm}^{-3}$, $f$ is of the order of a few
$10^{-3}\,\gamma_e$. The background dielectric constant $\epsilon_\infty$
takes
into account the effect of all the other non-resonant transitions and for an
atomic gas can be taken to be $1$ to a very good approximation.

\begin{figure}[htbp]
\includegraphics[width=3.3in]{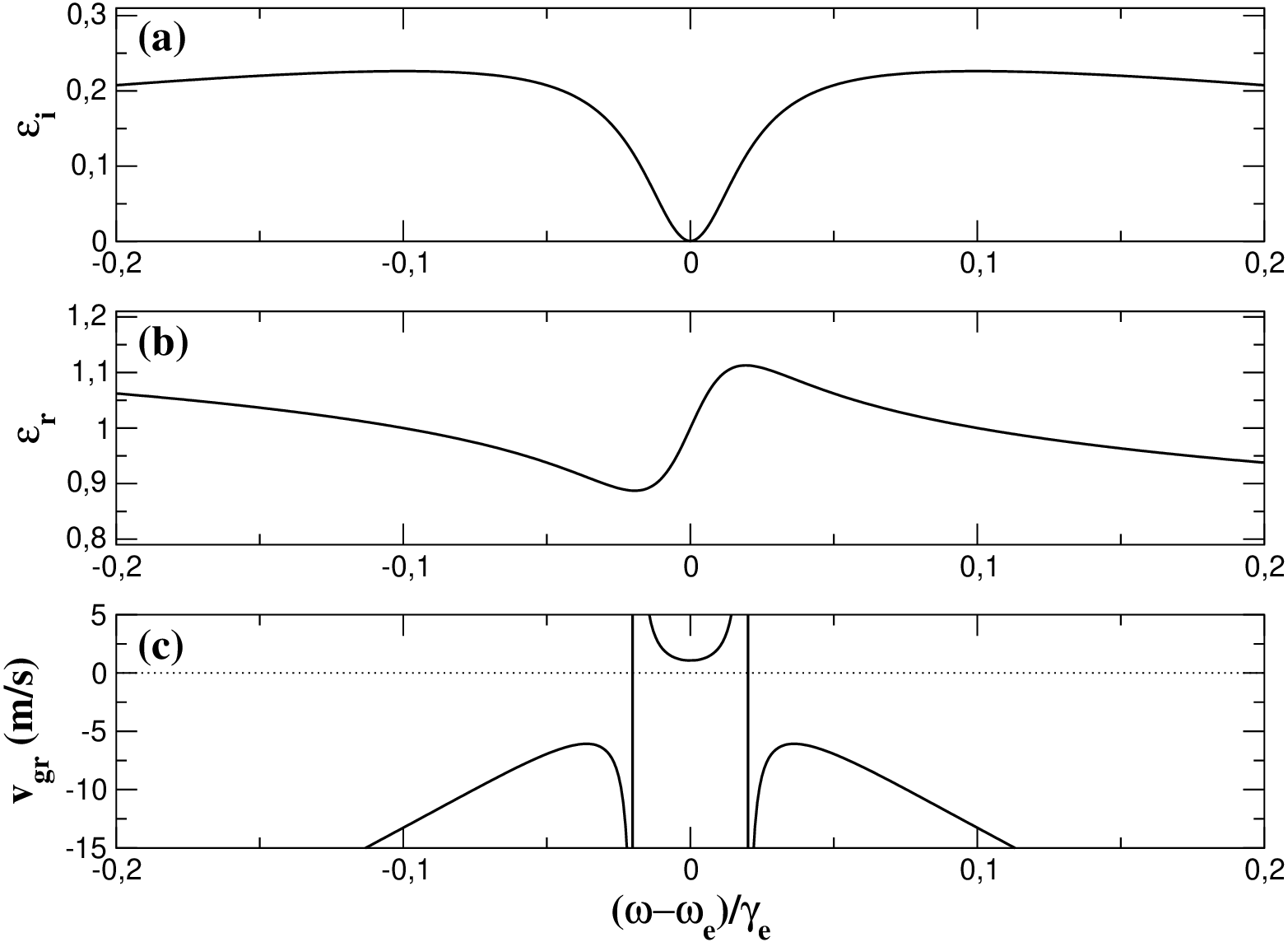}
\caption{Resonantly dressed EIT medium at rest: plot of the imaginary (a)
and real (b) parts of the dielectric function $\epsilon$ and of the
group velocity $v_{\rm gr}$ (c).
Medium parameters correspond to the case of an ultra cold $^{23}\textrm{Na}$
gas: $\gamma_e\simeq2\pi\cdot10\,\textrm{MHz}$,
$\lambda_e=589\,\textrm{nm}$, $f=0.009\,\gamma_e$, $\epsilon_\infty=1$,
$\gamma_m=10^{-4}\,\gamma_e$.  The Rabi frequency of the resonant
($\delta_c=0$) coupling beam is $\Omega_c=0.1\,\gamma_e$. For this choice of
parameters, the minimum (positive) group velocity is $v_{\rm
gr}\approx 1\,\textrm{m/s}$.
\label{fig:EpsRes}}
\end{figure}

For a resonantly dressed medium,
 i.e. $\delta_c=\omega_c-(\omega_e-\omega_m)=0$,
 the dielectric function
\eq{EpsEIT} in the neighborhood of the
resonance $|\omega-\omega_e|\ll\gamma_e$ can be rewritten as:
\begin{equation}
  \label{eq:EpsEITRes}
  \eps(\omega)=\epsilon_\infty+\frac{8\pi f i}{\gamma_e}\left[1+\frac{2i
\Omega_c^2/\gamma_e}{\omega_e-\omega-\frac{i}{2}(\gamma_m+\frac{4\Omega_c^2
}{\gamma_e})}\right].
\end{equation}
In the imaginary part of this expression shown in fig.\ref{fig:EpsRes}a, 
it is easy to seize a narrow dip around $\omega=\omega_e$
in the otherwise wide absorption profile of the $g\rightarrow e$ transition.
Provided $\Omega_c^2/\gamma_e\gg\gamma_m$, absorption
at the center of the dip is strongly suppressed, yielding nearly perfect
transparency for a resonant probe beam; this effect is the so-called {\em
  electromagnetically induced transparency} 
(EIT) effect~\cite{Alzetta}.
The linewidth of the dip is $\Gamma\simeq4\Omega_c^2/\gamma_e$ and
becomes strongly sub-natural ($\Gamma\ll\gamma_e$) for
$\Omega_c\ll\gamma_e$. The assumed
inequality $\gamma_m\ll\gamma_e$ guarantees that a good level of transparency
can be obtained simultaneously with a sub-natural linewidth of the dip.

In the dip region, 
the real part (fig.\ref{fig:EpsRes}b) of the dielectric function \eq{EpsEIT} shows an extremely steep
dispersion  yielding the group velocity:
\begin{equation}
\frac{v_{\rm gr}}{c}\simeq \frac{|\Omega_c|^2}{2\pi f\omega_e}
\eqname{Vg}
\end{equation}
which can be orders of magnitude slower than the vacuum value $c$
(fig.\ref{fig:EpsRes}c).
This suggests that the importance of the transverse Fresnel-Fizeau drag effect
\eq{Deltax} should be strongly enhanced in an EIT medium.

\begin{figure}[htbp]
\includegraphics[width=3.3in,clip]{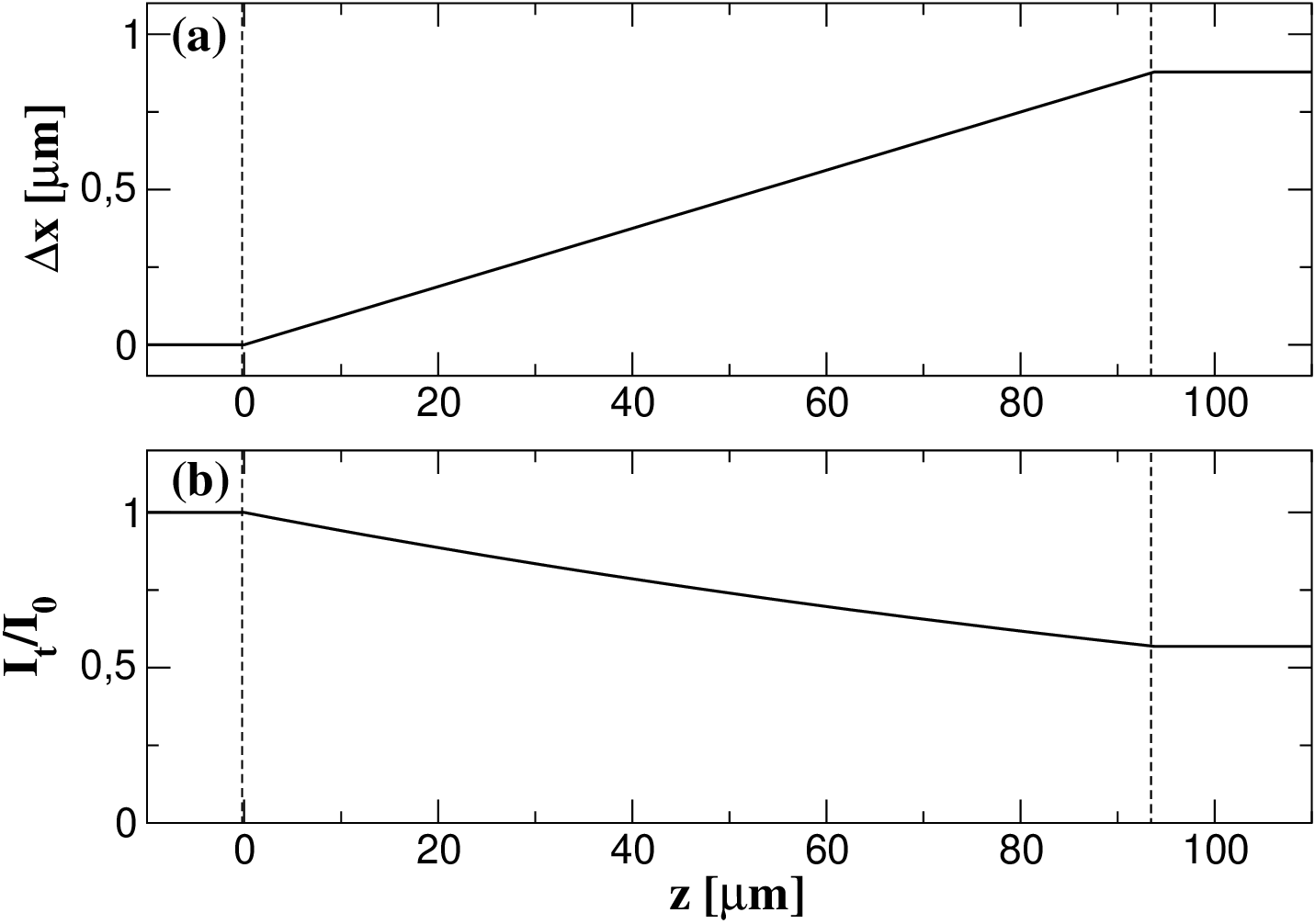}
\caption{Propagation of a light beam through a slowly moving
($v=0.01\,\textrm{m/s}$), resonantly dressed EIT medium. The optical 
parameters of the medium are the
same as in fig.\ref{fig:EpsRes}, the
vertical dashed lines correspond to the surfaces of the medium, whose thickness is
taken as $L=93\,\mu\textrm{m}$.
The incident beam is resonant
$\omega_0=\omega_e$ and its waist is $\sigma_0=20\,\mu\textrm{m}$.
{\em Downstream} transverse Fresnel-Fizeau drag (a) and corresponding weak absorption
of the beam (b).
The $x$ axis is oriented in the downstream direction.
\label{fig:dragRes}}
\end{figure}

To verify this prediction,
a complete calculation of the profile of the transmitted probe beam
can be numerically performed by
inserting the explicit expression of the dielectric function
\eq{EpsEIT} into the dispersion law \eq{FresnelWave} and then
performing the inverse Fourier transform \eq{InvFourier}.
For a probe exactly on resonance with the $g\rightarrow e$ transition
($\omega_0=\omega_e$), the numerical result plotted in fig.\ref{fig:dragRes}a
is in perfect agreement with the analytic prediction \eq{Deltax} when
the value \eq{Vg} for the group velocity is used.
The near absence of absorption at the center of the dip
guarantees that only a small fraction of the incident probe intensity is
absorbed by the medium (fig.\ref{fig:dragRes}b). 

Group velocities as low as $1\,\textrm{m/s}$ have been observed in
ultra cold atomic gases. For the 
choice of parameters ($v_{\rm gr}=1\,\textrm{m/s}$ and $L\approx
100\,\mu\textrm{m}$) made in fig.\ref{fig:dragRes}, a medium velocity of
the order of $v=0.01\,\textrm{m/s}$ gives a lateral shift of the order
of $\Delta x\approx
1\,\mu\textrm{m}$, orders of magnitude larger than the one originally
observed by Jones~\cite{Jones,Jones2}.
An even larger lateral shift $\Delta x$ should be obtained by
using a solid state material as dragging medium. 
Group velocities as slow
as $v_{\rm gr}=45\,\textrm{m/s}$ have in fact been recently observed in
a $\textrm{Pr}$
doped $\textrm{Y}_2\textrm{SiO}_5$~\cite{SlowLightSolids1} and
Ruby~\cite{SlowLightSolids2} crystals.  
Furthermore, the mechanical rigidity and the possibility of working at
higher temperatures should allow one to study the drag effect on a
thicker sample moving at a higher speed $v$.

\section{Anomalous transverse Fresnel-Fizeau drag effect}
\label{sec:Upstream}
The discussion of the previous sections has focused on the most
common case of media with {\em normal} dispersion.
As Kramers-Kronig causality relations~\cite{LandauECM} ensure that
\begin{equation}
\frac{c}{v_{\rm gr}}-\frac{v_{\rm ph}}{c}> 0,
\eqname{NormalDisp}
\end{equation}
for all frequency regions at which the medium is transparent and
non-amplifying, the corresponding transverse Fresnel-Fizeau drag \eq{Deltax}
results directed in the {\em downstream} direction, as if the light were to be
dragged by the moving medium.

On the other hand, several papers during the 70's~\cite{KoChuang,Lerche} have
discussed the possibility of having an {\em upstream}
transverse Fresnel-Fizeau drag in the presence of {\em anomalous} dispersion, i.e. 
in the presence of negative group velocity.
As negative group velocities in non-magnetic media are forbidden
by Kramers-Kronig relations~\cite{LandauECM}
 in all frequency regions where the medium is
transparent and non-amplifying, it has been possible to observe negative group
velocities only in the
presence of substantial
absorption~\cite{NegativeVgAbs-chu,NegativeVgAbs-steinberg,NegativeVgAbs-balcou}
or in amplifying 
media~\cite{NegativeVgAmpl}. Notice that the negative
group velocity effects which can be observed in left-handed media even in the
absence of absorption arise from a completely different mechanism, that is from a
simultaneously negative value of both the dielectric constant
$\epsilon$ and the magnetic susceptibility $\mu$~\cite{LeftHanded}.
This class of effects are excluded from the present discussion, which is
limited to non-magnetic ($\mu=1$) materials. 

In all experiments performed up to now, negative group velocities have been 
demonstrated by observing that the pulse advances in time with respect to the
same wave packet propagating in vacuum.
This kind of apparently superluminal behavior obviously refers to the peak
of the  wave packet only, and not to the propagation velocity of information;
as discussed in the review~\cite{MilonniReview}, the velocity of the front of
a step-function
signal can never exceed $c$.
\begin{figure}[htbp]
\includegraphics[width=3.3in]{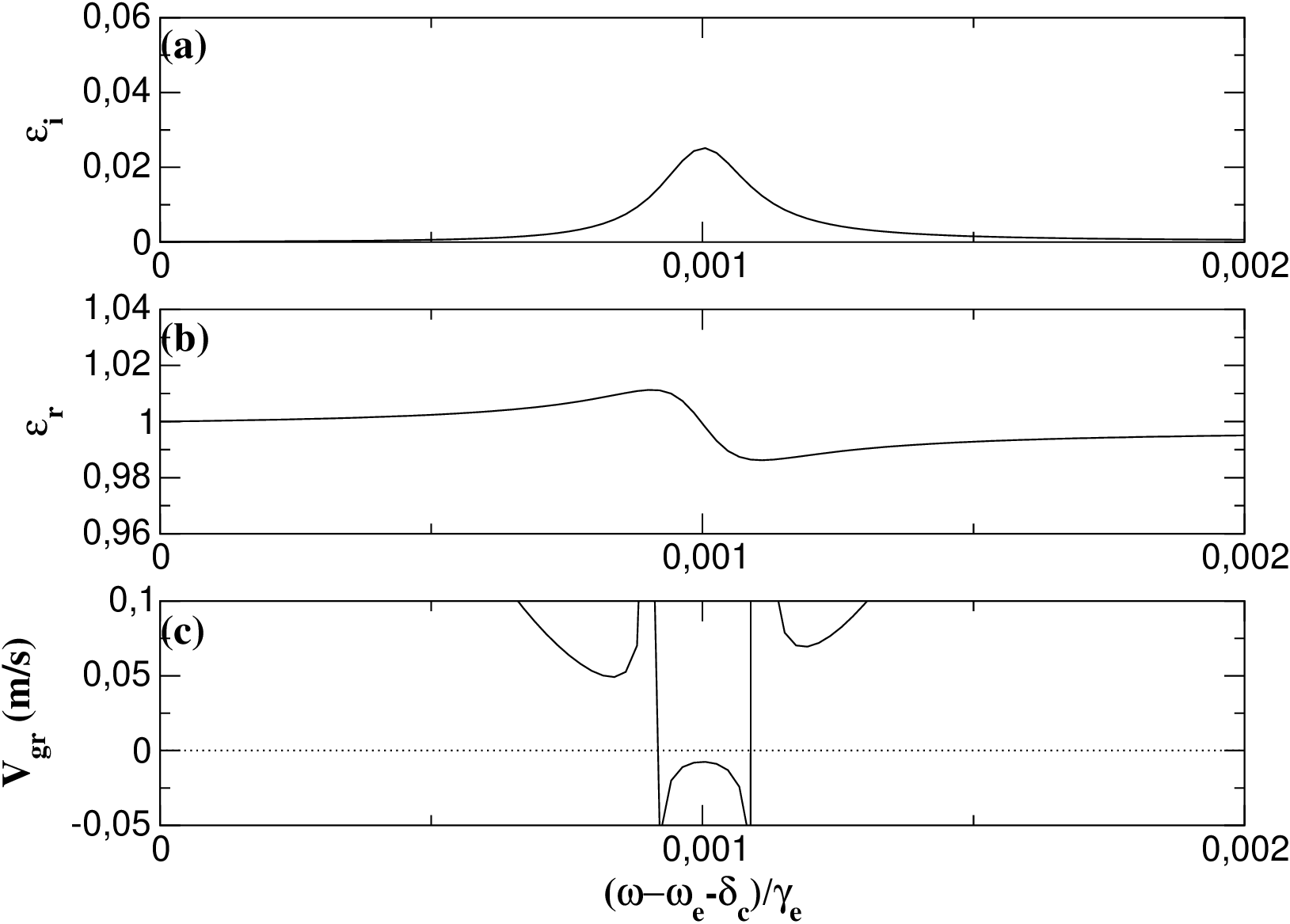}
\caption{Plot of the imaginary part $\epsilon_i(\omega)$ of the
dielectric function (a), of its real part $\epsilon_r(\omega)$
(b), and of the group velocity $v_{\rm gr}(\omega)$ (c) for
a non-resonantly ($\delta_c=10\,\gamma_e$) dressed EIT medium at rest.
The other parameters are the same as in fig.\ref{fig:EpsRes}.
\label{fig:EpsNRes}}
\end{figure}

As we have recently anticipated in~\cite{AnomFresTr}, a significant upstream
Fresnel-Fizeau drag effect should be observable in non-resonantly dressed EIT
media, when the coupling beam is not exactly on resonance with the
$m\rightarrow e$ transition, but has a finite detuning
$|\delta_c|=|\omega_m+\omega_c-\omega_e| \geq\gamma_e$.
This kind of media have in fact been shown to be good candidates for the
observation of ultra slow and negative group velocities~\cite{ArtoniNegVg}.
In the case $|\delta_c|\gg\gamma_e$, the dielectric function \eq{EpsEIT} in
the neighborhood of $\omega=\omega_m+\omega_c$
(fig.\ref{fig:EpsNRes}) can be written in the Lorentzian form:
\begin{equation}
  \label{eq:EpsEITNonRes}
  \eps(\omega)=\epsilon_\infty-\frac{4\pi
f}{\delta_c}+\frac{4\pi\,f\,\Omega_c^2}{\delta_c^2}\,\frac{1}
{\omega_2-\omega-\frac{i}{2}(\gamma_m+\frac{\Omega_c^2\gamma_e}{\delta_c^2})}.
\end{equation}
Largest absorptions occur at the Raman resonance with the two-photon transition
from the
ground $g$ state to the metastable $m$ state via the excited $e$ state, i.e.
at the two-photon resonance frequency
\begin{equation}
\eqname{omega2}
\omega_2=\omega_m+\omega_c+\frac{\Omega_c^2}{\delta_c}.
\end{equation}
The shift $\Omega_c^2/\delta_c$ from the bare resonant frequency
is due to the optical Stark effect induced by the coupling beam.
The linewidth of the resonance line is the sum of the bare linewidth of the
metastable $m$ level plus a contribution which takes into account its decay via 
the excited $e$ state:
\begin{equation}
  \label{eq:gamma2}
  \gamma_2=\gamma_m+\frac{\Omega_c^2}{\delta_c^2}\gamma_e.
\end{equation}
For large coupling beam detunings $|\delta_c|\gg\Omega_c$,
the linewidth $\gamma_2$ results much smaller than the natural linewidth of
the excited state, while the oscillator strength of the transition is itself
weakened:
\begin{equation}
  \label{eq:fosc2}
  f_2=\frac{\Omega_c^2}{\delta_c^2}f.
\end{equation}
The peak absorption (proportional to $f_2/\gamma_2$) does not vary
for increasing values of $|\delta_c|/\Omega_c$, at least
as far as $\gamma_2\gg\gamma_m$.
On the other hand, the anomalous dispersion at resonance $\omega=\omega_2$ is
under the same  conditions enhanced due to the narrower linewidth
$\gamma_2$.
The corresponding group velocity is given by
\begin{equation}
\frac{v_{\rm gr}}{c}=-\frac{\delta_c^2}{8\pi
  f\,\omega_e\,\Omega_c^2}
\left(\gamma_m+\frac{\Omega_c^2}{\delta_c^2}\,\gamma_e\right)^2
\eqname{NegVg}
\end{equation}
which, in the limit $\gamma_2\gg\gamma_m$, is a factor
$\gamma_e^2/4\delta_c^2$ slower in magnitude than the one obtained in the
resonant $\delta_c=0$ case considered in the previous section.
This means that the use of a detuned coupling with $|\delta_c|>\gamma_e$
should enable one to observe significant upstream drags over an optical thickness
$L$ of the order  of the absorption length, so that the transmitted probe
intensity still remains an
appreciable  fraction of the  incident one.

\begin{figure}[htbp]
\includegraphics[width=3.3in,clip]{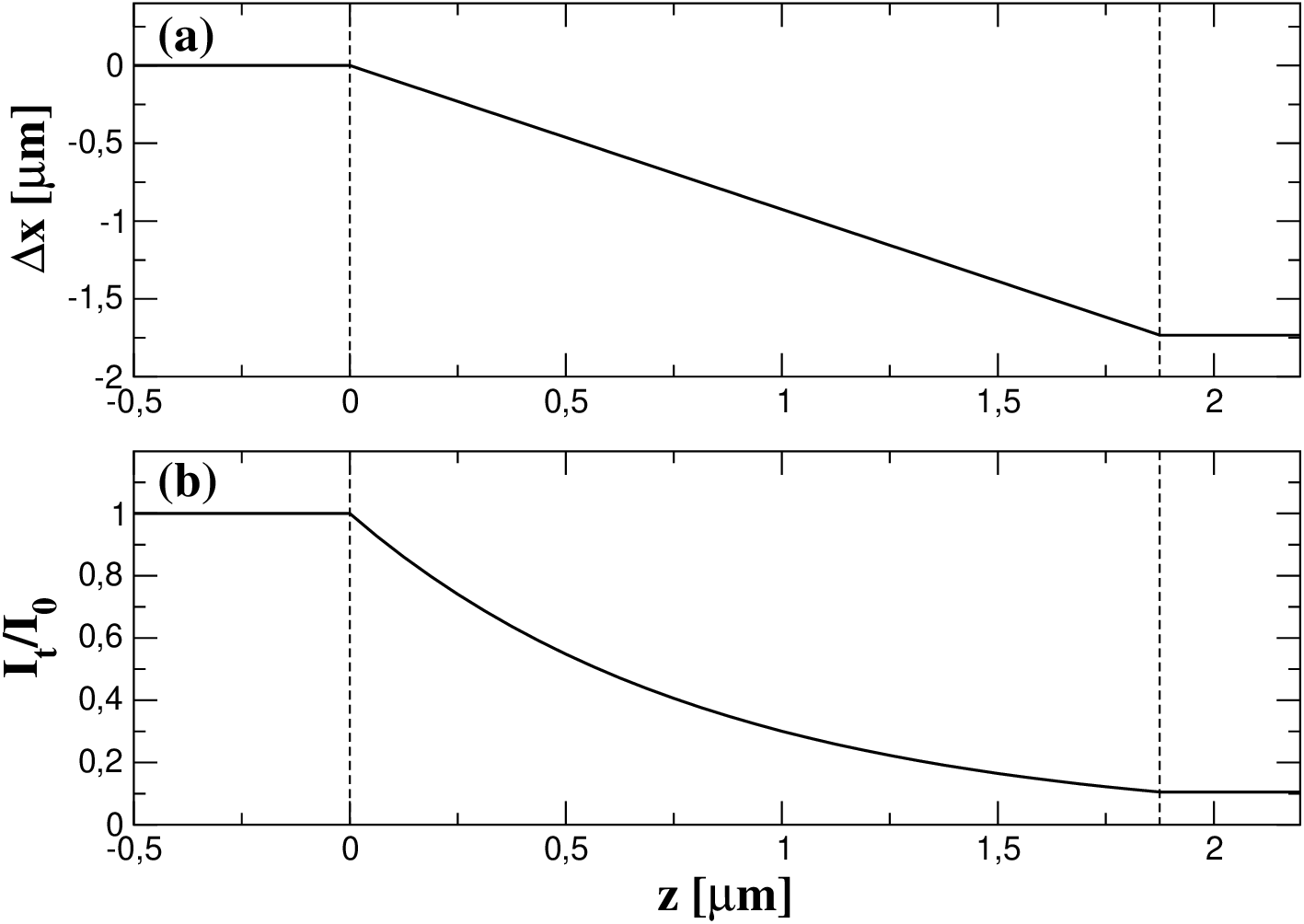}
\caption{Propagation of a light beam through a slowly moving
($v=0.01\,\textrm{m/s}$), non-resonantly dressed
($\delta_c=10\,\gamma_e$) EIT medium;
probe frequency on resonance with the two-photon transition
($\omega_0=\omega_2$).
The optical parameters of the medium are the same as in fig.\ref{fig:EpsNRes};
the 
vertical dashed lines correspond to the surfaces of the medium, whose thickness is
taken as $L=1.9\,\mu\textrm{m}$.
The incident
beam waist is $\sigma_0=20\,\mu\textrm{m}$.
Anomalous {\em upstream} transverse Fresnel-Fizeau drag (a) and corresponding
absorption of the beam (b). The $x$ axis is oriented in the downstream
direction. \label{fig:dragNRes}}
\end{figure}

In order to verify this expectation, we have again inserted
the explicit expression of the dielectric function \eq{EpsEIT} into the
propagation equation \eq{FresnelWave} and we have numerically performed the
inverse Fourier transform so as to obtain the profile of the transmitted beam
(fig.\ref{fig:dragNRes}a).
As expected, light bendings are found to occur 
in the upstream direction and the magnitude of the effect is in
good agreement with the approximated analytical expression \eq{Deltax} in
which the imaginary part of $\epsilon$ was neglected.
We have also verified in fig.\ref{fig:dragNRes}b that the beam
is not completely absorbed during propagation through the medium.
Notice that for the same set of material parameters and the same experimental 
configuration as in the previous section, except for the coupling beam
detuning $\delta_c$, the group velocity $|v_{\rm gr}|$ is
now of the order of
$0.01\,\textrm{m/s}$, i.e. a factor $100$ slower than in the case of a
resonant coupling. This explains the bigger deflection
angle $\theta$.

Non-resonantly dressed EIT media are therefore good candidates for the
experimental observation of significant upstream deflections and
consequently large anomalous transverse Fresnel-Fizeau drags.
Such an observation should finally resolve a controversy started in the
seventies about the possibility of such an effect~\cite{KoChuang,Lerche}.
Furthermore, the anomalous upstream Fresnel-Fizeau drag could also provide an
interesting way of demonstrating negative group velocities, in alternative to
the
usual~\cite{NegativeVgAbs-chu,NegativeVgAbs-steinberg,NegativeVgAbs-balcou,NegativeVgAmpl}  
measurement of the negative temporal
delay of the pulse after its propagation across the medium.

\section{Higher order dispersion effects}
\label{sec:HigherOrder}
The effects discussed in the previous sections
entirely rely on the strongly reduced value of the group velocity of light in
EIT media. 
For slow enough medium velocities, higher order
dispersion effects such as the dispersion of absorption or the group
velocity  dispersion are indeed very small and can be hardly observed in the
results plotted in figs.\ref{fig:dragRes} and
\ref{fig:dragNRes}. 
On the other hand, these terms may be no longer negligible for sufficiently
large values of the medium velocity, regime in which the light propagation can 
exhibit qualitatively new features
\footnote{Even at the highest velocities here considered ($v$ of
  the order of a few $\textrm{m/s}$), the use
  of the linearized form of the Lorentz transformations
  (\ref{eq:Lorentz1}-\ref{eq:Lorentz3}) is well justified: as the
  magnitude of the transverse Doppler effect of the probe and coupling beam
  $|\Delta \omega_\perp|\simeq \omega_0 v^2 / c^2 \lesssim
10^{-8}\,\gamma_e$ the
  contribution of the terms in $v^2$ arising from the relativistic $\gamma$
  factor is negligible as compared to the one coming from the dispersion of
  $\epsilon$.}.
In the following of this section we shall discuss in detail two examples of
such effects.

\subsection{Light propagation along a curved path}
\label{sec:Tilt}

In the present subsection, we shall discuss the effect of non-rectilinear
light propagation due to simultaneously large dispersions of both 
absorption~\cite{RL-1} (proportional to $d\epsilon_i/d\omega$) and
group velocity (proportional to $d^2\epsilon_r/d\omega^2$).

For large enough values of the absorption dispersion
$d\epsilon_i/d\omega$, the absorption coefficient can have 
significant variations across the range of transverse $k_x$ vectors
present in the incident beam.
In this case, the center of mass of the $k_x$ wave-vector distribution
\begin{equation}
  \label{eq:spectrum_kx}
  k_x^{\rm cm}(z)=\frac{\int\!dk_x\,dk_y\,k_x\,|{\tilde E}(k_x,k_y;z)|^2}{
\int\!dk_x\,dk_y\,|{\tilde E}(k_x,k_y;z)|^2}
\end{equation}
 moves from its initial value $k_x^{\rm cm}(z=0)=0$ as
 the  beam  propagates through the medium.
For a spatially wide Gaussian incident beam, we can limit the expansion of
 $\epsilon_i$ to the linear terms in $k_{x,y}$ and we find that the
 distribution
 of transverse wave vectors keeps a gaussian  shape, but its center of mass 
$k_x^{\rm cm}(z)$ shifts to: 
 \begin{equation}
  \label{eq:spectrum_kx_z}
  k_x^{\rm cm}(z)=-\frac{z}{\sigma_0^2}\frac{\partial
  \textrm{Im}[k_z\al{in}]}{\partial k_x}
\end{equation}
For a negative value of $\partial \textrm{Im}[k\al{in}_z]/\partial k_x$, the
$k_x>0$ components result in fact less absorbed than the $k_x<0$ ones so that
the 
Gaussian spectrum shifts towards the $k_x^{\rm cm}>0$ region;
vice versa for a positive value of $\partial \textrm{Im}[k\al{in}_z]/\partial
k_x$. 

In the slow group velocity regime ($v_{\rm gr}'\ll c$), the dependence of the
propagation wavevector \eq{k_zeps} on the transverse wave vector $k_{x,y}$
mainly comes from the Doppler effect combined with the strong frequency
dispersion of the dielectric function, so that the transmission phase reads as:
\begin{equation}
  \label{eq:Phi_slow}
  \Phi(k_x,k_y;z)\simeq\frac{\omega_0\,z}{c}\sqrt{\epsilon(\omega_0-k_x v)}.
\end{equation}
By applying the same stationary-phase arguments used in sec.\ref{sec:Downstream}
to the finite $k_x^{\rm cm}(z)$ case,
we find that after propagation for a
distance $z$ the spatial center of mass of the beam is located at
\begin{equation}
  \label{eq:transv_sh_disp}
  \Delta x(z)=z\frac{v}{v'_{\rm gr}(\omega_0-v\, k_x^{\rm cm}(z)\,)}.
\end{equation}

In physical terms, as the spectral center of mass $k_x^{\rm cm}$ of the
beam varies with $z$ 
because of the filtering action of the absorption, the transverse
Fresnel-Fizeau drag
at a given position $z$  has to be evaluated using the group
velocity $v'_{\rm gr}$ at the Doppler shifted frequency $\omega_0-v\, k_x^{\rm
  cm}(z)$. 
In the presence of group velocity dispersion, the variation
of 
$k_x^{\rm cm}(z)$ with $z$ implies that the light beam is no longer
rectilinear, but acquires a finite curvature.
Notice in particular that the analytical expression for the
angle $\theta(z)$ between the trajectory \eq{transv_sh_disp} and the normal
$z$ direction
\begin{multline}
  \label{eq:tangent}
  \tan \theta(z)=\frac{\partial \Delta x(z)}{\partial z}= \\
=\frac{v}{v'_{\rm gr}(\omega_0-vk_x^{\rm cm})}+\frac{v^2\,z}{\big[v'_{\rm
gr}(\omega-vk_x^{\rm cm})\big]^2}\,\frac{dv'_{\rm
gr}}{d\omega}\,\frac{dk_x^{\rm cm}}{dz}
\end{multline}
now contains a term explicitly depending on the spectral center of mass shift
  $dk_x^{\rm  cm}/dz$ which was not present in \eq{theta}.

A related effect was discussed in the time domain
in~\cite{MuschiettiDum} where  a small time-dependence of the
group velocity for a light pulse propagating through a dispersive dielectric
was predicted to show up as a consequence of the combined 
  effect of the frequency dispersions of absorption and of group
  velocity. 

\begin{figure}[htbp]
\includegraphics[width=3.3in,clip]{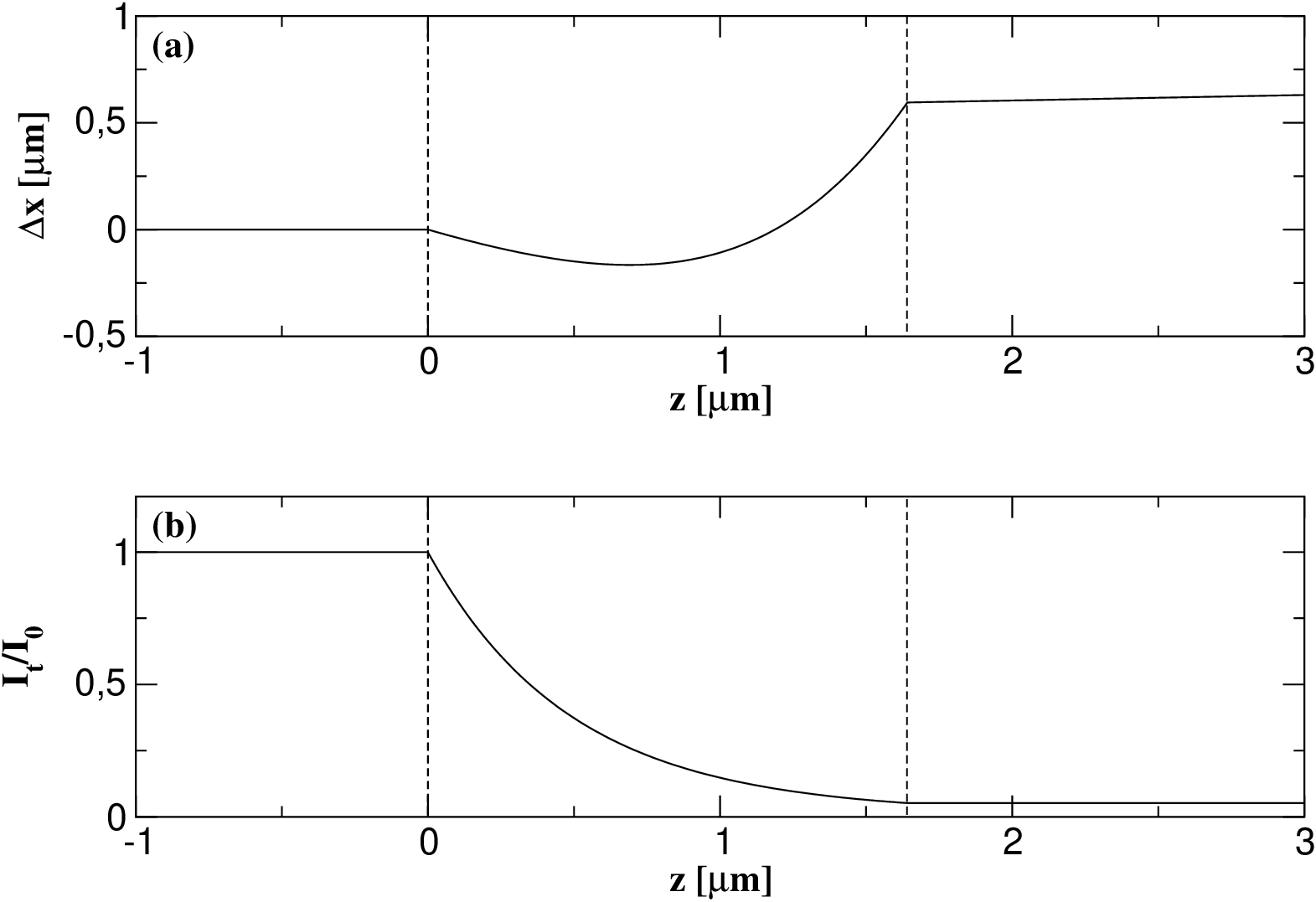}
\caption{Non-rectilinear light beam propagation through a moving
($v=4\,\textrm{m/s}$) EIT medium for a resonant coupling ($\delta_c=0$) and
a slightly detuned probe ($\omega_0-\omega_e=0.04\,\gamma_e$); the
incident beam waist is $\sigma_0=3\,\mu\textrm{m}$.
(a) panel: curved beam path across the moving medium. (b) panel:
intensity of the light beam at different depths in the medium.
The optical parameters of the medium are the same as in fig.\ref{fig:EpsRes};
the vertical dashed lines correspond to the surfaces of the medium, whose
thickness is
taken as $L=1.65\,\mu\textrm{m}$.
\label{fig:tilt1}}
\end{figure}

\begin{figure}[htbp]
\includegraphics[width=3.3in,clip]{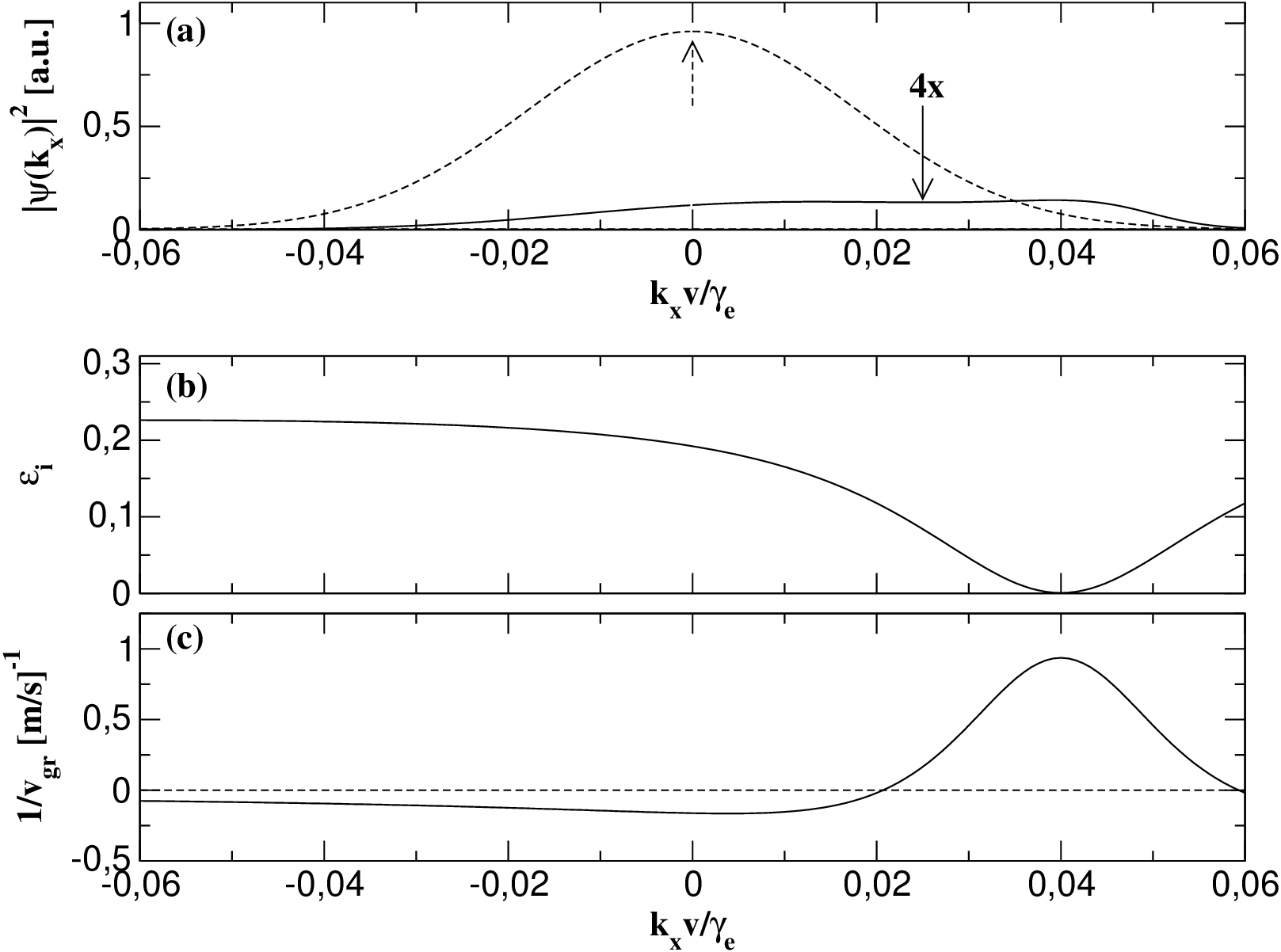}
\caption{
Physical interpretation of the non-rectilinear light propagation of
fig.\ref{fig:tilt1}. (a) panel: spatial
Fourier transform of the incident (dashed) and transmitted (solid)
beam profile; for each spectrum, the arrow indicates the position of the
spectral center of mass. For the sake of clarity, the transmitted spectrum has
been multiplied by 4.
(b) and (c) panels: absorption and inverse group velocity $1/v_{\rm gr}$
spectra as a function of the transverse wavevector $k_x$. The corresponding
Doppler-shifted frequency in the rest frame $\Sigma'$ is $\omega'=\omega_0-k_x
v$. Since $\omega_0-\omega_e=0.04\, \gamma_e$, the resonance is found at
$k_x v=0.04\, \gamma_e$.
\label{fig:tilt2}}
\end{figure}

Figs.\ref{fig:tilt1}-\ref{fig:tilt2}, report the result of numerical
calculations for the specific case of a moving EIT medium when the coupling
beam is resonant ($\delta_c=0$) and the probe beam is slightly
detuned towards the blue ($\omega_0-\omega_e=0.04\,\gamma_e$).
As one can verify by comparing the spectrum in fig.\ref{fig:tilt2}a with the
group velocity spectrum in fig.\ref{fig:tilt2}c, most of the incident beam
spectrum lies in the negative group velocity region ($k_x v<0.02\,
\gamma_e$), so the beam is initially dragged in the upstream direction.
As absorption is weaker for the $k_x>0$ components for which the
Doppler-shifted frequency 
$\omega'=\omega_0-k_x v$ is closer to resonance ($\partial
  \textrm{Im}[k_z\al{in}]/\partial k_x<0$, see fig.\ref{fig:tilt2}b), the
  negative $k_x<0$ 
components   are more rapidly quenched than the positive $k_x>0$ components
and the center of mass $k_x\al{\rm cm}$ of the spectral distribution moves towards the positive
$k_x>0$ values.
As one can see in fig.\ref{fig:tilt2}c, $\frac{\partial}{\partial
  k_x}\frac{1}{v'_{\rm gr}}>0$ in the region
$0<k_x v<0.04\, \gamma_e$ of interest and therefore the
curvature of the beam will
be towards the downstream direction.
After the first $1\,\mu\textrm{m}$ of propagation, most of the $k_x$
spectrum is found  in the positive $v_{\rm gr}'$ region
($k_x v>0.02\, \gamma_e$), so that the transverse Fresnel-Fizeau drag is from
now on in the downstream direction (fig.\ref{fig:tilt1}a).

As a consequence of the shift of the spectral center of mass $k_x^{\rm
  cm}$, the beams exits from the rear face of the medium at a small
but finite angle with respect to the normal towards the downstream direction.
Although this effect is hardly visible on the scale of fig.\ref{fig:tilt2}a,
this small bending of the beam direction may have a significative effect
on the subsequent rectilinear propagation of the beam in the free space.

Unfortunately, this effect of non-rectilinear propagation is associated to a
  rather severe absorption of the beam; for the specific case in figure,
  the transmitted intensity is of the order of 5\% of the incident one
  (fig.\ref{fig:tilt1}b).

It is also worth noticing that the curvature effect described in the present section follows from
  a reshaping of the beam in momentum space and hence is physically
  different from the ones discussed in~\cite{ulf-prl-2000,ulf-pra-2000}, which 
  instead originate from a non-uniform velocity field of the slow-light
medium.

\subsection{Temporal and spatial coherence restoration}

If both probe and coupling beams are exactly on resonance
($\omega_0-\omega_e=\delta_c=0$), both absorption $d\eps_i/d\omega'$ and
group velocity $d^2\eps_r/d{\omega'}^2$ dispersion vanish, so that
the effects described in sec.\ref{sec:Tilt} do not take place.
In the present subsection, we shall show how one can rather take advantage of
the large value of $d^2\eps_i/d{\omega'}^2$ to improve the coherence
level of a noisy incident probe beam. In sec.\ref{sec:Temporal}, the
case of {\em temporal} coherence restoration in a stationary EIT medium will
be adressed, while in sec.\ref{sec:Spatial} we shall show how the same
concepts can be applied in the case of a moving EIT medium to improve
the level of {\em spatial} coherence.

\subsubsection{Temporal coherence restoration}
\label{sec:Temporal}

Consider an incident probe beam of carrier frequency $\omega_0$ whose
complex amplitude $E_0(t)$ is assumed as fluctuating in time over a
characteristic time scale $\tau_c$: 
\begin{equation}
E(t)=E_0(t)\,e^{-i\omega_0 t}.
\end{equation}
Following a standard model~\cite{Loudon}, the decay in time 
of the first-order coherence function $g\al{1}(\tau)$ is taken to be Gaussian:
\begin{equation}
g\al{1}(\tau)=\frac{\langle E_0^*(\tau)\,E_0(0)\rangle}
{\langle E_0^*(0)\,E_0(0)\rangle}
=\exp(-\tau^2/2\tau_c^2).
\end{equation}
The frequency spectrum $|{\tilde  E}(\omega)|^2$ of the beam, being
  proportional to the Fourier transform of the coherence function
  $g\al{1}(\tau)$, is also Gaussian with linewidth $\sigma_c=1/\tau_c$.

In a one-dimensional geometry, the propagation of a pulse
through a stationary dielectric medium is described by the usual
Fresnel equation:
\begin{equation}
k_z^2(\omega)=\epsilon(\omega)\,\frac{\omega^2}{c^2}.
\end{equation}
In the neighborhood of the resonance at $\omega_e$, the dielectric
function of a slow light EIT medium can be approximately written as:
\begin{equation}
\epsilon(\omega)\simeq 1+\frac{2c}{\omega_e v_{\rm
gr}}(\omega-\omega_e)+i\frac{\alpha}{2}(\omega-\omega_e)^2
\end{equation}
where $v_{\rm gr}$ is the group velocity at resonance and the real quantity
$\alpha$ is given by:
\begin{equation}
  \label{eq:alpha}
  \alpha=\left.\frac{d^2\epsilon_i(\omega)}{d\omega^2}\right|
_{\omega=\omega_e}=  \frac{4\pi f \gamma_e}{\Omega_c^4}.
\end{equation}
At the lowest order in $\omega-\omega_e$, the real and imaginary parts of the
wavevector $k_z(\omega)$ can then be written as:
\begin{eqnarray}
\left.k_z(\omega)\right|_r&\simeq&\frac{\omega_e}{c}\Big[1+\frac{c}{\omega_e
  v_{\rm gr}}(\omega-\omega_e)\Big]. \\
\left.k_z(\omega)\right|_i&\simeq&\frac{\alpha\omega_e}{4c}(\omega-\omega_e)^2
\end{eqnarray}
In particular, notice how the imaginary part of $k_z$ is proportional
to the square of $\omega-\omega_e$. Since after propagation over a distance
$L$ the amplitude of each frequency component is multiplied by a factor
$\exp[i k_z(\omega)L]$, the frequency spectrum after propagation keeps its Gaussian
shape, but the frequency linewidth is reduced to: 
\begin{equation}
\sigma_c(L)=\frac{\sigma_c}{\sqrt{1+\frac{\alpha\sigma_c^2\omega_e}{c}L}}
\eqname{sigma_c}
\end{equation}
and the coherence time correspondingly increased to:
\begin{equation}
\tau_c(z)=\tau_c\sqrt{1+\frac{\alpha\omega_e}{c \tau_c^2}L}.
\end{equation}
Since the EIT medium does not provide amplification, a drawback of this
filtering technique is that part of the incident intensity is lost during the
line-narrowing process; the beam intensity after propagation through a
distance $L$ is in fact equal to:
\begin{equation}
I(L)=\frac{\sigma_c(L)}{\sigma_c(0)}I(0).
\end{equation}

\subsubsection{Spatial coherence restoration}
\label{sec:Spatial}

For a moving EIT medium and a strictly monochromatic light beam at $\omega_0$,
a similar effect occurs in the $k_x$ space.
As discussed in full detail in sec.\ref{sec:GeneralTh}, the
propagation of light is described in this case by equation \eq{k_zeps}.
For resonant probe and coupling beams, the imaginary part of
$\textrm{Im}[k_z\al{in}]$ corresponding to absorption is
proportional to $v^2 k_x^2$:
\begin{equation}
\textrm{Im}[k_z\al{in}]\simeq \frac{\alpha\omega_0}{4c}v^2 k_x^2.
\end{equation}
As the amplitude of the transverse spatial fluctuations is
proportional to the amplitude of non-vanishing $k_x$ components, the spatial
profile of the beam flattens 
as the beam propagates through the moving EIT medium. The faster the speed $v$
of the medium, the more efficient the spatial coherence restoration process.

\begin{figure}[htbp]
\includegraphics[width=3.3in]{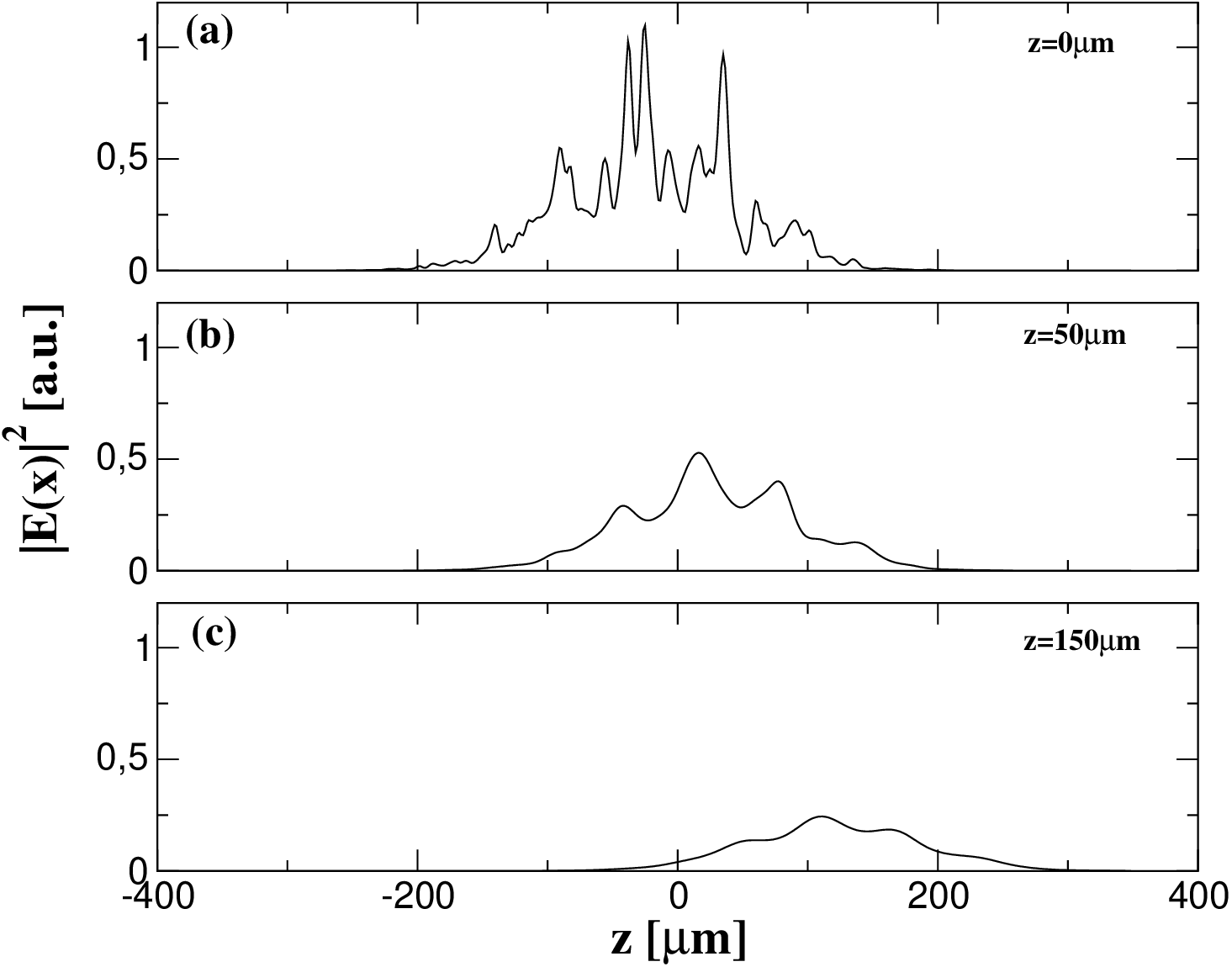}
\caption{Transverse spatial coherence restoration during propagation
across a moving ($v=1\,\textrm{m/s}$) EIT medium in a fully resonant regime
($\delta_c=\omega_0-\omega_e=0$, $\omega_0=\omega_e$). (a) panel: noisy
incident beam profile. (b,c) panels: beam profile after propagation
through the medium.
Same medium parameters as in fig.\ref{fig:EpsRes}-\ref{fig:tilt1}.
\label{fig:filter}}
\end{figure}

Results of a numerical calculation for a spatially noisy incident beam
propagating through a moving EIT medium are presented in fig.\ref{fig:filter}.
Notice how the fluctuation amplitude is strongly suppressed
during propagation. At the end of the process, the losses in the total
intensity amount to 60\%.

The overall shift of the beam which can be seen in the figure is
due to the transverse Fresnel-Fizeau drag effect discussed in detail in
sec.\ref{sec:DownstreamEIT}. Since both the probe and the coupling are
resonant, the shift is directed in the downstream direction.

\section{Conclusions}
\label{sec:Conclu}

In the present paper we have given a comprehensive analysis of the transverse
Fresnel-Fizeau drag effect for light
propagating across a uniform slab of moving EIT medium. 
All calculations have been
performed using realistic parameters taken from
EIT experiments with ultra-cold atomic clouds. 
Since our results are
essentially a consequence of electromagnetically induced transparency,
they are extremely general and thus hold through
also for the recently prepared solid state EIT media.
Depending on the detuning of the probe and
coupling beams with respect to the medium resonance, different
regimes have been identified. 

In the presence of a slow and positive group velocity, the magnitude of the
downstream Fresnel-Fizeau drag effect is predicted
to be significantly enhanced with respect to previous experiments. In the
regime of negative group velocity, an
anomalous upstream Fresnel-Fizeau drag has been predicted to occur. Not only
would 
this help in solving a long-standing
controversy on the observability of such an effect, but it could also
provide an interesting alternative way for
experimentally detecting a negative group velocity.

For larger values of the velocity of the moving medium, higher order
dispersion terms such as the dispersion of
absorption and the dispersion of group velocity have been shown to play an
important role in the phenomenology of
light propagation. Depending on the specific choice of probe and coupling
detuning, light propagation along curved
paths can be observed, as well as the restoration of spatial coherence of a
noisy beam.

The extension of the present analysis to more complicated geometries including
non-uniformly moving slow light media as well as probe pulses of finite
duration will be the subject of future studies.

\begin{acknowledgments}
One of us (M.A.) should like to thank U. Leonhardt for enlightening discussions on the issue of slow light in
moving media. Financial support from the EU (Contracts HPMF-CT-2000-00901 and
HPRICT1999-00111), from the INFM (project PRA "photonmatter") and from the MIUR 
(grant PRIN 2002-028858) is greatly acknowledged. 
Laboratoire Kastler Brossel is a unit\'e de Recherche de l'Ecole normale
sup\'erieure et de l'Universit\'e Pierre et Marie Curie, associ\'ee au CNRS.

\end{acknowledgments}


\begin{thebibliography}{99}

\bibitem{SlowLightColdAtoms-hau} L. V. Hau,
S. E. Harris, Z. Dutton, and C. H. Behroozi, Nature {\bf 397}, 594 (1999);

\bibitem{SlowLightColdAtoms-inouye}S. Inouye {\em et al.}, Phys. Rev. Lett. {\bf 85}, 4225 (2000).

\bibitem{SlowLightHotAtoms-kash} M. M. Kash {\em et al.}, Phys. Rev. Lett.  {\bf 82}, 5229
(1999);

\bibitem{SlowLightHotAtoms-budker}
D. Budker , D. F. Kimball, S. M. Rochester, and V. V. Yashchuk, Phys. Rev. Lett. {\bf 83}, 1767 (1999).

\bibitem{SlowLightSolids1} A. V. Turukhin {\em et al.},
Phys. Rev. Lett. {\bf 88}, 023602 (2002).

\bibitem{SlowLightSolids2} M. S. Bigelow, N. N. Lepeshkin, and
R. W. Boyd, Phys. Rev. Lett. {\bf 90}, 113903 (2003).

\bibitem{LuceFermaTh} O. Kocharovskaya, Y.
Rostovtsev, and M. O. Scully, Phys. Rev. Lett. {\bf 86}, 628
(2001).

\bibitem{LuceFermaExp-phillips} D.  F.  Phillips {\em et al.},
Phys. Rev. Lett. {\bf 86}, 783 (2001);

\bibitem{LuceFermaExp-hau}
C. Liu, Z. Dutton, C. H. Behroozi, and L. V. Hau, Nature, {\bf 409}, 490, (2001).

\bibitem{Alzetta} G. Alzetta, A. Gozzini, L. Moi, and G. Orriols, Nuovo
  Cimento {\bf 36B}, 5 (1976); E. Arimondo, G. Orriols, Lett. Nuovo Cimento
\textbf{17}, 333 (1976).


\bibitem{EITReviewArimondo}  E. Arimondo in {\em Progress in Optics XXXV},
ed. by E.
Wolf, Elsevier Science, (1996) pag.257.

\bibitem{Hansch} T.W. H\"ansch, P.E. Toschek, Z. Phys. {\bf 236}, 213 (1970).

\bibitem{HarrisReview} S. Harris, Phys. Today {\bf 50}, No. 7, 36 (1997).

\bibitem{SlowLightReview} A. B. Matsko {\em et al.},
Adv. At. Mol. Opt. Phys. {\bf 46}, 191 (2001).

\bibitem{Marangos} J. Marangos, J. Mod. Opt. \textbf{45}, 471 (1998).


\bibitem{NLO-kasapi} S. E. Harris, J. E. Field, and A. Kasapi, Phys. Rev. A {\bf 46},
  R29 (1992);

\bibitem{NLO-harris}S. E. Harris and L. V. Hau, Phys. Rev. Lett. {\bf 82}, 4611,
(1999);

\bibitem{NLO-lukin}
M. D. Lukin and A. Imamoglu, Phys. Rev. Lett. {\bf 84}, 1419 (2000).

\bibitem{AtomEntangl}  M. D. Lukin, S. F. Yelin, and M. Fleischhauer,
Phys. Rev. Lett. {\bf 84}, 4232 (2000).

\bibitem{QuantumMem} M. Fleischhauer and
M. D. Lukin, Phys. Rev. Lett. {\bf 84}, 5094 (2000).

\bibitem{AcoustoOpt}  A. B. Matsko,
Y. V. Rostovtsev, H. Z. Cummins, and M. O. Scully,
Phys. Rev. Lett. {\bf 84}, 5752 (2000).


\bibitem{ulf-prl-2000}
U. Leonhardt and P. Piwnicki, Phys. Rev. Lett. {\bf 84}, 822 (2000);

\bibitem{ulf-pra-2000}U. Leonhardt, Phys.
Rev. A {\bf 62}, 012111 (2000);

\bibitem{ulf-jmo-2001}U. Leonhardt and P. Piwnicki, J. Mod. Opt. {\bf 48}, 977 (2001);

\bibitem{ulf-black-hole}U. Leonhardt, Phys.
Rev. A {\bf 65}, 043818 (2002);

\bibitem{ulf-nature} U.Leonhardt, Nature {\bf 415}, 406 (2000);

\bibitem{ulf-pulse}J. Fiurasek, U. Leonhardt and R. Parentani, Phys.
Rev. A {\bf 65}, 011802 (2002);

\bibitem{ulf-physicsworld}
U. Leonhardt, Phys. Rev. A {\bf 62}, 012111 (2000). Physics World, \textbf{15}, No
2, 7, (2002)


\bibitem{ArtoniNegVg} M. Artoni, G. C. La Rocca, F. S. Cataliotti, and
F. Bassani, Phys. Rev. A {\bf 63}, 023805 (2001).


\bibitem{FresLongEIT}
M. Artoni, I. Carusotto, G. C. La
Rocca, and F. Bassani, Phys. Rev. Lett. {\bf 86}, 2549 (2001).


\bibitem{Jones} R. V. Jones, Proc. Roy. Soc. London A {\bf 328},
337 (1972).

\bibitem{Jones2} R. V. Jones, Proc. Roy. Soc. London A {\bf 345}, 351 (1975).


\bibitem{KoChuang} H. C. Ko and C. W. Chuang, Astroph. Journ. {\bf
222}, 1012 (1978); H. C. Ko, Astroph. Journ. {\bf 231}, 589 (1979).

\bibitem{Lerche} I. Lerche, Astroph. Journ. {\bf 187}, 589 (1974).

\bibitem{Jackson} J. D. Jackson, {\em Classical
Electrodynamics} (J. Wiley, New York, 1975).


\bibitem{FresTrTh-player}  M. A. Player, Proc. Roy. Soc. London A {\bf 345},
343 (1972).

\bibitem{FresTrTh-rogers} G. L. Rogers,  Proc. Roy. Soc. London A {\bf 345},
345 (1972).

\bibitem{AnomFresTr} M. Artoni, I. Carusotto, G. C. La Rocca, and F. Bassani,
  J. Opt. B {\bf 4},  S345 (2002).


\bibitem{LandauECM} L. D. Landau and E. M. Lifshitz, {\em Electrodynamics
of continuous media}, Pergamon Press, London, 1960.


\bibitem{NegativeVgAbs-chu} S. Chu and S. Wong, Phys. Rev. Lett. {\bf 48}, 738
  (1982);

\bibitem{NegativeVgAbs-steinberg}A. M. Steinberg, P. G. Kwiat, and R. Y. Chiao, Phys. Rev. Lett. {\bf
    71}, 701 (1993);

\bibitem{NegativeVgAbs-balcou}Ph. Balcou and L. Dutriaux, Phys. Rev. Lett. {\bf 78},
  851 (1997).


\bibitem{NegativeVgAmpl} L. J. Wang, A. Kuzmich, and A. Dogariu, Nature {\bf
    406}, 277 (2000).

\bibitem{LeftHanded} V.G.Veselago, Sov. Phys. Usp. {\bf 10}, 509 (1968);
D.R.Smith, {\em et al.}, Phys. Rev. Lett. {\bf 84},  4184 (2000)

\bibitem{MilonniReview} P. W. Milonni, J. Phys. B: At. Mol. Opt. Phys. {\bf
    35}, R31 (2002).





\bibitem{RL-1} M. Artoni and R.Loudon, Phys. Rev. A {\bf 55}, 1347 (1997).

\bibitem{MuschiettiDum} L. Muschietti and C. T. Dum, Phys. Fluids B
{\bf 5}, 1383 (1993).


\bibitem{Loudon} R. Loudon, {\em The quantum theory of light}, Clarendon
  Press, Oxford, 1973.








\end{thebibliography}
\end{document}